\documentclass[11pt,twoside]{article}

\usepackage{fancyhead}
\usepackage{psfrag}
\usepackage{epsfig}
\usepackage{graphicx}
\usepackage{amsmath}
\usepackage{amssymb}
\usepackage{amsthm}
\usepackage{subfigure}
\usepackage{cite}
\usepackage[utf8]{inputenc}
\usepackage{color}
\usepackage{url}
\usepackage{tikz}
\usepackage{pgfplots}
\usepackage[english]{babel}
\usepackage{algorithm}
\usepackage{algorithmic}
\usepackage{trfsigns}
\newcommand{\fra}{\mathcal{A}}
\newcommand{\frb}{\mathcal{B}}
\newcommand{\frr}{\mathcal{RM}}

\newcommand{\FF}{\mathbb{F}}

\DeclareMathOperator{\dH}{d_H}

\begin{document}

\vspace*{5mm}

\noindent
%%--The title should be placed here
\textbf{\LARGE Error Correction for Physical Unclonable Functions Using Generalized Concatenated Codes
%%--If you do not use any support, just comment the next line or delete it.
%\footnote{This research is partially supported by ...}
}
\thispagestyle{fancyplain} \setlength\partopsep {0pt} \flushbottom
\date{}

\vspace*{5mm}
\noindent
%%--The name(s) of the author(s), their e-mails and addresses should be placed here
\textsc{Sven M\"uelich} \hfill \texttt{sven.mueelich@uni-ulm.de} \\
\textsc{Sven Puchinger} \hfill \texttt{sven.puchinger@uni-ulm.de} \\
\textsc{Martin Bossert} \hfill \texttt{martin.bossert@uni-ulm.de} \\
{\small Institute of Communications Engineering, University of Ulm, Germany} \\
\textsc{Matthias Hiller} \hfill \texttt{matthias.hiller@tum.de} \\
\textsc{Georg Sigl} \hfill \texttt{sigl@tum.de} \\
{\small Institute for Security in Information Technology, TU M\"unchen, Germany}

\medskip

\begin{center}
\parbox{11,8cm}{\footnotesize
%%--The abstract goes here.
\textbf{Abstract.} Physical Unclonable Functions can be used for secure key generation in cryptographic applications. It is explained how methods from coding theory must be applied in order to ensure reliable key regeneration. Based on previous work, we show ways how to obtain better results with respect to error probability and codeword length. Also, an example based on Generalized Concatenated codes is given, which improves upon used coding schemes for PUFs.
}
\end{center}

\baselineskip=0.9\normalbaselineskip

% ===========================================================================
% Introduction
% ===========================================================================

\section{Introduction}
Two of the most challenging tasks when developing a cryptosystem are to implement secure key generation and storage \cite{Lenstra2012,Torrance2009}. Keys have to be random, unique and unpredictable. %Since keys have to be used by algorithms, it is necessary that they are stored in memory after they were generated. Both, badly implemented key generators and unprotected key storage enable physical attacks to adversaries. Physically Unclonable Functions (PUFs) can be used in order to tackle these two problems. 
\emph{Physical Unclonable Functions} (PUF) possess an intrinsic randomness due to process variations during manufacturing. In addition, the key is also implicitly stored in the PUF. However, the results when regenerating a key vary, which can be interpreted as errors. Thus, error correction must be used in order to compensate this effect.
%Because their intrinsic randomness which results from process variations during manufacturing is much stronger than the randomness of many key generator implementations, PUFs can be used for secure key generation. Since a PUF can be used to re-generate the same key multiple times, it is not necessary to store the key in memory. However, the results when generating one and the same key multiple times can vary. This problem can be avoided by applying error correcting codes.

Previous work on this topic used standard constructions, e.g. a concatenated scheme of a BCH and Repetition code in \cite{MaesCryptoPaper2012}, which can be improved. The intention of this paper is to propose better code constructions for the application in PUFs. The main requirements for the codes are low decoding complexity and high error correction capability.

In the remainder of the paper we first explain PUF basics and how coding theory is used for key generation using PUFs. Section~\ref{sec_constructions} describes methods and codes suitable for this scenario. Finally, an example code construction, improving those commonly used for PUFs, is given in Section~\ref{sec_examples}. The last section concludes the paper.

% ===========================================================================
% Physically Unclonable Functions
% ===========================================================================
\newpage
\section{Physical Unclonable Functions}
\label{sec_pufs}
\cite{Dodis2008} describes a PUF as a physical entity which uses an input (challenge) in order to produce an output (response), where a challenge can result in different responses when applied to a certain PUF instance several times. The distance of two such responses is called \emph{intra-distance}\footnote{With distance we mean the \emph{Hamming} distance $\dH$.}. Reasons for these varying responses are random noise, measurement uncertainties, aging and changing environmental conditions like temperature or supply voltage. A small response intra-distance is preferred, since in applications there is a need for reproducibility of responses. Similarly, we can apply the same challenge to two different PUF instances and call the distance of the responses \emph{inter-distance}, which results from variations during the manufacturing process. This measure gives us the distinguishability of different PUF instances, which is preferred to be large. Unclonable means the hardness of manufacturing a PUF with a specific challenge-response-behavior.
There are many possibilities to realize PUFs, e.g. delay-based (e.g. Ring Oscillator PUFs) or memory-based (e.g. SRAM PUFs). An overview of popular types can be found in \cite{MaesDiss2012}.

PUFs can be used in order to realize secure key generation and storage for cryptographic applications.
Due to the static randomness over the PUFs lifetime, it is possible to regenerate a key repeatedly instead of storing it permanently.
As described above, PUF responses are not exactly reproducible and therefore a response cannot be used as key directly. Hence, methods of coding theory must be used to circumvent this drawback. A common way to do this is \emph{Secure Sketching} \cite{Dodis2008}. When a key is generated the first time using a PUF response $y$, a \emph{Sketch} function is used to extract and store \emph{helper data} $\omega$ of $y$. A possible realization of such a function can use the syndrome of a linear code as helper data, i.e. $\omega = H y^{T}$, where $H$ is a parity check matrix of the code.
Since the syndrome contains only information about the difference to the next codeword, an attacker knowing the helper data is left with an uncertainty as large as the number of codewords.
For regenerating a key, the same challenge must be used to obtain a response $y'$ which is likely to differ from $y$. 
%This can be interpreted as an additive error $y'=y+e$, where for most PUFs the error model can be interpreted as a binary symmetric channel (BSC) with crossover probability $p$, where $p$ is given by the PUF. 
For most PUFs, this can be interpreted as an additive error $y'=y+e$, resulting from a binary symmetric channel (BSC) with crossover probability $p$, where $p$ is given by the PUF. 
If $\dH\left(y,y'\right)$ is small enough, a \emph{Recover} function (cf. Algorithm \ref{alg_recover}) is able to recover the key $y$ using $y'$ and $\omega$.

\begin{algorithm}
\caption{Recover}          
\label{alg_recover}                          
\begin{algorithmic}
    \REQUIRE $y' = y+e$ and $\omega = H y^{T}$
    \STATE $s = H y'^{T}-w = H e^{T}$
    \STATE Use decoder to obtain $e$ from $s$.
    \RETURN $y'' = y' - e$
\end{algorithmic}
\end{algorithm}

There are different possibilities for realizing a Secure Sketch based on error-correcting codes, for example the syndrome construction presented above. Other possibilities are the Code-Offset Construction \cite{MaesDiss2012}, Index-Based Syndrome coding \cite{Yu2010}, Complementary Index-Based Syndrome coding \cite{Hiller2012} and Differential Sequence Coding \cite{Hiller2013}.

The main challenge is to find good codes that can be used for Secure Sketches.
Since PUF responses are not uniformly distributed, the generated keys are often hashed by a cryptographic hash function before they are used. The combination of a Secure Sketch and a hash function is usually referred to as \emph{Fuzzy Extractor}. % (cf. \cite{dodis2008fuzzy}).

% ===========================================================================
% Code Constructions
% ===========================================================================

\section{Code Constructions}
\label{sec_constructions}

We first give constraints one has to deal with when designing codes for PUFs. Since decoding is usually implemented on a hardware device, there are constraints regarding time and area optimization. Also, the designed codes must be binary. A typical constraint is to design a code which has a block error probability $P_{err}$ smaller than a certain threshold for a given BSC crossover probability $p$. The dimension of the code must be greater than or equal to the length of the key that should be generated. The length of the codewords can be chosen arbitrarily, but one has to keep in mind that for generating one key, as many bits as the codeword length have to be extracted from the PUF. In this section, we describe construction and decoding methods which are suitable for this scenario.

\subsection{Reed--Muller Codes}
\label{subsec_rm}

A \emph{Reed--Muller} (RM) code $\frr(r,m)$ of order $r$ with $r \leq m$ is a binary linear code with parameters $n = 2^{m}$, $k = \sum_{i = 0}^{r} \binom{m}{i}$ and $d=2^{m-r}$.
%Every Reed-Muller code can be constructed by generalized multiple concatenation of repetition codes, single parity check codes and a block code of length 2 \cite{SB94}.
RM codes work well for PUF Secure Sketching due to an easily implementable decoding, e.g. \cite{SB94}. \emph{Simplex} codes are RM codes with parameters $(1,m)$.

\subsection{Generalized Minimum Distance Decoding}
\label{subsec_GMD}

\emph{Generalized Minimum Distance} (GMD) decoding (cf. \cite{forney1966generalized}) is a method to increase the number of correctable errors beyond half the minimum distance by incrementally declaring the least reliable positions of a received word to be erasures. Hence, \emph{soft-information} and \emph{error-erasure} decoders are needed.

\subsection{Generalized Concatenated Codes}
\label{subsec_gcc}

The authors of \cite{Bosch2008} found that concatenated codes are advisable for implementing Secure Sketches.
Instead of ordinary concatenated codes, we propose using \emph{Generalized Concatenated} (GC) codes as introduced in \cite{Zyablov_Shavgulidze_Bossert_1999,Bossert1999}.
A GC code with given $n$ and $d$ contains more codewords and hence has a higher code rate than an ordinary concatenated code with the same parameters.
%Similarly, having the same length and dimension, a GCC has a larger minimum distance than an ordinary concatenated code.

The main idea of GC codes is to partition an inner code $\frb^{(1)}$ into multiple levels of subcodes.
Let $\frb^{(i)}_{j}$ denote the $j$-th subcode at partition level $i$.
The goal is to create partitions such that the minimum distances of the subcodes increase strictly monotonically from level to level in the partition tree. Each codeword of $\frb^{(1)}$ can be uniquely determined using a numeration of the partition. This numeration is protected by outer codes. Code $\fra^{(i)}$ denotes the outer code which protects the numeration of the partition from level $i$ to level $i+1$. For a detailed description of GC codes we refer to \cite{Bossert1999}.

% ===========================================================================
% Example
% ===========================================================================

\section{Example}
\label{sec_examples}

In \cite{MaesCryptoPaper2012}, a design for cryptographic key generators based on PUFs was introduced, using a concatenation of a $(318,174,35)$ BCH code and a $(7,1,7)$ Repetition code in order to generate a $128$ bit key with error probability $10^{-9}$.

In this section we want to give an example code construction for the same scenario which improves existing schemes in code length, block error probability and easiness of the implementation. As error model we choose a BSC with crossover probability $p = 0.14$. We want to generate a $128$ bit key. Thus, we have to choose a code with dimension $\geq 128$ and a block length less than the one used in \cite{MaesCryptoPaper2012}, namely $2226$. The block error probability $P_{err}$ should be less than $10^{-9}$.

We choose a generalized concatenation of an inner $(16,5,8)$ Simplex code $\frb^{(1)}$ and RM codes of length $128$ as outer codes $\fra^{(i)}$. Hence, we obtain a code of length $128 \cdot 16 = 2048$, i.e. it can be represented as a matrix with $128$ rows, each containing a codeword of the Simplex code.

The inner code $\frb^{(1)}$ is partitioned into $16$ disjoint subcodes $\frb^{(2)}_{i}$ with parameters $(16,1,16)$, e.g. $\frb^{(2)}_{0000}$ can be the repetition code of length $16$ and all other elements of the partition are its distinct cosets. The enumeration $i \in \{0000, \dots, 1111\}$ is then protected by four $\frr(128,8,64)$ codes, one for each bit. Since the subcodes $\frb^{(2)}_{i}$ contain exactly two elements each, we can again partition them into subcodes containing only one element, $\frb^{(2)}_{i,0}$ and $\frb^{(2)}_{i,1}$. The enumeration $\{0,1\}$ is then protected by a $\frr(128,99,8)$ code. Thus, we can encode $4 \cdot 8 + 99 = 131 \geq 128$ bits.

We decode in two steps. First, we apply \emph{Maximum Likelihood} (ML) decoding\footnote{This can be done easily since $\frb^{(1)}$ contains only $32$ codewords} of the $\frb^{(1)}$ code to each row of the codeword matrix. If we cannot decode uniquely, we declare this row to be an erasure. Otherwise, we decode (i.e. save the information word of length $5$) and also save the Hamming distance of the received column to the decoding output as reliability information. After applying this to every of the $128$ rows, we obtain a $128 \times 5$ matrix containing either $5$ bits of information or erasures in each row. Then, we apply GMD decoding of the $\frr(128,8,64)$ code, using the reliability information given by the number of errors corrected before, to each of the first four columns of the matrix. If decoding does not fail, we obtain the correct enumeration $i \in \FF_2^4$ of the subcode $\frb^{(2)}_{i}$ for each row.

Thus, for each row, we know in which subcode $\frb^{(2)}_{i}$ we have to decode in the second step. Since $\frb^{(2)}_{i}$ contains only two elements, we can again apply ML decoding and obtain one bit for each row of the matrix, including soft information like in the first step, or an erasure. Then we can apply GMD decoding of the $\frr(128,99,8)$ code to get the fifth column of the "information matrix".

Besides theoretical analysis, simulations have shown that for a BSC with $p=0.14$, the block error probability is reduced to $5.37 \cdot 10^{-10}$. We also decreased the codeword length from $2226$ to $2048$. Another advantage of our construction is that decoding is easier to implement, since we only use codes with decoders working over $\FF_2$. Table~\ref{tab:summary} summarizes the improvements.

\begin{table}[h]
\renewcommand{\arraystretch}{1.5}
\centering
\begin{tabular}{p{1.5cm}|p{2.5cm}|p{2.5cm}|p{2.5cm}}
Code
& $P_{err}$
& Length
& Largest Field\footnotemark \\
\hline \hline
\cite{MaesCryptoPaper2012}
& $\approx 10^{-9}$
& $2226$
& $\FF_{2^8}$ (BCH) \\
\hline
New
& $\approx 5.37 \cdot 10^{-10}$
& $2048$
& $\FF_2$ \\
\end{tabular}
\label{tab:summary}
\caption{Improvements of the new code construction compared to \cite{MaesCryptoPaper2012}.}
\end{table}
\footnotetext{Largest field used by decoder. Operations over small fields are usually easier to implement.}

% ===========================================================================
% Conclusion and Further Results
% ===========================================================================

\section{Conclusion and Future Work}
\label{sec_conclusion}

We explained how coding theory is used for regenerating cryptographic keys using PUFs. Moreover, we proposed code constructions and decoding methods which improve existing coding schemes for PUFs and illustrated these by giving an example. RM codes turn out to be efficient in this scenario. However, their rates cannot be chosen arbitrarily, which would improve the design opportunities for GC codes. Hence, for example generalized concatenation using \emph{Reed--Solomon} codes as outer codes can be used. In addition, more methods from coding theory can be examined for suitability in this setting.

\newpage

\bibliographystyle{IEEEtran}
\bibliography{puf_gcc}

\begin{thebibliography}{10}
\providecommand{\url}[1]{#1}
\csname url@samestyle\endcsname
\providecommand{\newblock}{\relax}
\providecommand{\bibinfo}[2]{#2}
\providecommand{\BIBentrySTDinterwordspacing}{\spaceskip=0pt\relax}
\providecommand{\BIBentryALTinterwordstretchfactor}{4}
\providecommand{\BIBentryALTinterwordspacing}{\spaceskip=\fontdimen2\font plus
\BIBentryALTinterwordstretchfactor\fontdimen3\font minus
  \fontdimen4\font\relax}
\providecommand{\BIBforeignlanguage}[2]{{%
\expandafter\ifx\csname l@#1\endcsname\relax
\typeout{** WARNING: IEEEtran.bst: No hyphenation pattern has been}%
\typeout{** loaded for the language `#1'. Using the pattern for}%
\typeout{** the default language instead.}%
\else
\language=\csname l@#1\endcsname
\fi
#2}}
\providecommand{\BIBdecl}{\relax}
\BIBdecl

\bibitem{Lenstra2012}
A.~K. Lenstra, J.~P. Hughes, M.~Augier, J.~W. Bos, T.~Kleinjung, and
  C.~Wachter, ``Ron was wrong, whit is right,'' Cryptology ePrint Archive,
  Report 2012/064, 2012.

\bibitem{Torrance2009}
R.~Torrance and D.~James, ``{The State-of-the-Art in IC Reverse Engineering},''
  in \emph{CHES}, 2009, pp. 363--381.

\bibitem{MaesCryptoPaper2012}
R.~Maes, A.~Herrewege, and I.~Verbauwhede, ``{PUFKY: A Fully Functional
  PUF-Based Cryptographic Key Generator},'' in \emph{CHES}, 2012, pp. 302--319.

\bibitem{Dodis2008}
Y.~Dodis, R.~Ostrovsky, L.~Reyzin, and A.~Smith, ``{Fuzzy Extractors: How to
  Generate Strong Keys from Biometrics and Other Noisy Data},'' \emph{SIAM
  Journal on Computing}, vol.~38, no.~1, pp. 97--139, Mar. 2008.

\bibitem{MaesDiss2012}
R.~Maes, ``{Physically Unclonable Functions: Constructions, Properties and
  Applications},'' Ph.D. dissertation, KU Leuven, 2012.

\bibitem{Yu2010}
M.-D.~M. Yu and S.~Devadas, ``{Secure and Robust Error Correction for Physical
  Unclonable Functions},'' \emph{IEEE Design \& Test of Computers}, vol.~27,
  no.~1, pp. 48--65, 2010.

\bibitem{Hiller2012}
M.~Hiller, D.~Merli, F.~Stumpf, and G.~Sigl, ``{Complementary IBS: Application
  Specific Error Correction for PUFs},'' in \emph{IEEE Int. Symp. on
  Hardware-Oriented Security and Trust}, 2012.

\bibitem{Hiller2013}
M.~Hiller, M.~Weiner, L.~Rodrigues~Lima, M.~Birkner, and G.~Sigl, ``{Breaking
  Through Fixed PUF Block Limitations with Differential Sequence Coding and
  Convolutional Codes},'' in \emph{TrustED}, 2013, pp. 43--54.

\bibitem{SB94}
G.~Schnabel and M.~Bossert, ``{Reed Muller Codes as Generalized Multiple
  Concatenated Codes with Soft-Decision Decoding},'' \emph{Internal Report,
  Informationstechnik, University of Ulm, Germany}, 1994.

\bibitem{forney1966generalized}
G.~Forney~Jr, ``{Generalized Minimum Distance Decoding},'' \emph{IEEE
  Transactions on Information Theory}, vol.~12, no.~2, pp. 125--131, 1966.

\bibitem{Bosch2008}
C.~B\"{o}sch, J.~Guajardo, A.-R. Sadeghi, J.~Shokrollahi, and P.~Tuyls,
  ``{Efficient Helper Data Key Extractor on FPGAs},'' in \emph{CHES}, 2008, pp.
  181--197.

\bibitem{Zyablov_Shavgulidze_Bossert_1999}
V.~Zyablov, S.~Shavgulidze, and M.~Bossert, ``\BIBforeignlanguage{en}{{An
  Introduction to Generalized Concatenated Codes}},''
  \emph{\BIBforeignlanguage{en}{European Transactions on Telecommunications}},
  vol.~10, no.~6, p. 609–622, 1999.

\bibitem{Bossert1999}
M.~Bossert, \emph{{Channel Coding for Telecommunications}}, 1st~ed.\hskip 1em
  plus 0.5em minus 0.4em\relax New York, NY, USA: John Wiley \& Sons, Inc.,
  1999.

\end{thebibliography}

\end{document}